\documentclass[12pt]{article}
\usepackage[dvips]{graphicx}

\newcommand{\ie}{{\it i.e.}}
\newcommand{\eg}{{\it e.g.}}
\renewcommand{\bar}[1]{\overline{#1}}
\newcommand{\half}  {\frac{1}{2}}

\newcommand{\qu}{{\rm q}}
\newcommand{\qb}{${\rm\bar q}$}

\newcommand{\pvec}{\vec p}
\newcommand{\kvec}{\vec k}
\newcommand{\rvec}{\vec r}
\newcommand{\Rvec}{\vec R}

\newcommand{\ieps}{i\varepsilon}

\newcommand{\pl}{{||}}

\newcommand{\order}[1]{${ O}\left(#1 \right)$}

\newcommand{\eq}[1]{(\ref{#1})}

\newcommand{\beq}{\begin{equation}}
\newcommand{\eeq}{\end{equation}}
\newcommand{\beqa}{\begin{eqnarray}}
\newcommand{\eeqa}{\end{eqnarray}}
\newcommand{\VEV}[1]{\left\langle{#1}\right\rangle}
\newcommand{\ket}[1]{\vert{#1}\rangle}

\begin {document}
\begin{flushright}
{\small
SLAC--PUB--8998\\
September 2001\\}
\end{flushright}

\begin{center}
{{\bf\LARGE Diffraction Dissociation in QCD\ and \\[1ex] Light-Cone
Wavefunctions}\footnote{Work supported by Department of Energy
contract  DE--AC03--76SF00515.}}

\bigskip
{\it Stanley J. Brodsky \\
Stanford Linear Accelerator Center \\
Stanford University, Stanford, California 94309 \\
E-mail:  sjbth@slac.stanford.edu}
\medskip
\end{center}

\vfill

\begin{center}
{\bf\large
Abstract }
\end{center}

The diffractive dissociation of a hadron at high energies, by
either Coulomb or Pomeron exchange, has a natural description in
QCD as the materialization of the projectile's light-cone
wavefunctions; in particular, the diffractive dissociation of a
meson, baryon, or photon into high transverse momentum jets
measures the shape and other features of the projectile's
distribution amplitude.  Diffractive dissociation can thus test
fundamental properties of QCD, including color transparency and
intrinsic charm. All of these effects have an impact on exclusive
decays of $B$ mesons and the study of $CP$ violation. I also
discuss recent work which shows that the structure functions
measured in deep inelastic lepton scattering are affected by
final-state rescattering, thus modifying their connection to
light-cone probability distributions.  In particular, the
shadowing of nuclear structure functions is due to destructive
interference effects from leading-twist diffraction of the virtual
photon, physics not included in the nuclear light-cone
wavefunctions. \vfill

\begin{center}
{\it Presented at the Workshop on \\
 9th Blois Workshop On Elastic And Diffractive Scattering\\
 Pruhonice, Prague, Czech Republic\\
9-15 June 2001}\\
\end{center}
\vfill
\newpage

\section{Introduction}

The diffractive dissociation of a hadron at high energies, by
either Coulomb or Pomeron exchange, can be understood as the
materialization of the projectile's light-cone wavefunctions; in
particular, the diffractive dissociation of a meson, baryon, or
photon into high transverse momentum jets measures the shape and
other features of the projectile's distribution amplitude,
$\phi(x_i,Q),$ the valence wavefunction which controls high
momentum transfer exclusive amplitudes. Diffractive dissociation
can also test fundamental properties of QCD, including color
transparency and intrinsic charm.

Diffractive dissociation in QCD can be understood as a three-step
process:

1.  The initial hadron can be decomposed in terms of its quark and
gluon constituents in terms of its light-cone Fock-state
components.  For example, the eigensolution of a
negatively-charged meson QCD,  projected on its color-singlet $B =
0$, $Q = -1$, $J_z = 0$ eigenstates $\{\ket{n} \}$ of the free
Hamiltonian $ H^{QCD}_{LC}(g = 0)$ at fixed $\tau = t-z/c$ has the
expansion:
\begin{eqnarray}
\left\vert \Psi; P^+, {\vec P_\perp}, \lambda \right> &=& \sum_{n
\ge 2,\lambda_i} \int \Pi^{n}_{i=1} {d^2k_{\perp i} dx_i \over
\sqrt{x_i} 16 \pi^3}
 16 \pi^3 \delta\left(1- \sum^n_j x_j\right) \delta^{(2)}
\left(\sum^n_\ell \vec k_{\perp \ell}\right) \cr &&\left\vert n;
x_i P^+, x_i {\vec P_\perp} + {\vec k_{\perp i}}, \lambda_i\right
> \psi_{n/p}(x_i,{\vec k_{\perp i}},\lambda_i)
 . \nonumber
\end{eqnarray}
The light-cone momentum fractions $x_i = k^+_i/P^+_\pi = (k^0 +
k^z_i)/(P^0+P^z)$ with $\sum^n_{i=1} x_i = 1$ and ${\vec k_{\perp
i}}$ with $\sum^n_{i=1} {\vec k_{\perp i}} = {\vec 0_\perp}$
represent the relative momentum coordinates of the QCD
constituents independent of the total momentum of the state.  The
actual physical transverse momenta are ${\vec p_{\perp i}} = x_i
{\vec P_\perp} + {\vec k_{\perp i}}.$ The $\lambda_i$ label the
light-cone spin $S^z$ projections of the quarks and gluons along
the quantization $z$ direction.  The physical gluon polarization
vectors $\epsilon^\mu(k,\ \lambda = \pm 1)$ are specified in
light-cone gauge by the conditions $k \cdot \epsilon = 0,\ \eta
\cdot \epsilon = \epsilon^+ = 0.$ The parton degrees of freedom
are thus all physical; there are no ghost or negative metric
states.  Remarkably, the light-cone Fock wavefunctions
$\psi_{n/p}(x_i,\vec k_{\perp i},\lambda_i)$ are independent of
the proton's momentum $P^+ = P^0 + P^z$, and $P_\perp$.  The
wavefunctions represent the ensembles of states possible when the
hadron is intercepted by a light-front at fixed $\tau = t+z/c.$
The light-cone representation thus provide a frame-independent,
quantum-mechanical representation of the incoming hadron at the
amplitude level, capable of encoding its multi-quark, hidden-color
and gluon momentum, helicity, and flavor correlations in the form
of universal process-independent hadron wavefunctions.

Progress in measuring the basic parameters of electroweak
interactions and $CP$ violation will require a quantitative
understanding of the dynamics and phase structure of $B$ decays at
the amplitude level. The light-cone Fock representation is
specially advantageous in the study of exclusive $B$ decays.  For
example, Dae Sung Hwang~\cite{Brodsky:1999hn} and I have derived
an exact frame-independent representation of decay matrix elements
such as $B \to D \ell \bar \nu$ from the overlap of $n' = n$
parton-number conserving wavefunctions and the overlap of
wavefunctions with $n' = n-2$ from the annihilation of a
quark-antiquark pair in the initial wavefunction.

2.  In the second step, the incoming hadron is resolved by Pomeron
or Odderon (multi-gluon) exchange with the target or by Coulomb
dissociation. The exchanged interaction has to supply sufficient
momentum transfer $q^\mu$ to put the diffracted state $X$ on
shell.  Light-cone energy conservation requires $q^- = {( m_X^2-
m_\pi^2)/ P^+_\pi},$ where $m_X$ is the invariant mass of $X$.  In
a heavy target rest system, the longitudinal momentum transfer is
$q^z = {(m_X^2- m_\pi^2)/ E_{\pi {\rm lab}}}.$ Thus the momentum
transfer $t = q^2$ to the target can be sufficiently small so that
the target remains intact.

In perturbative QCD, the pomeron is generally be represented as
multiple gluon exchange between the target and projectile.
Effectively this interaction occurs over a short light-cone time
interval, and thus like photon exchange, the perturbative QCD
pomeron can be effectively represented as a local operator.  This
description is believed to be applicable when the pomeron has to
resolve compact states and is the basis for the terminology ``hard
pomeron".  The BFKL formalism generalizes the perturbative QCD
treatment by an all-orders perturbative resummation, generating a
pomeron with a fixed Regge intercept $\alpha_P(0)$.  Next to
leading order calculations with BLM scale fixing leads to a
predicted intercept $\alpha_P(0) \simeq
0.4$~\cite{Brodsky:1999kn}.   However, when the exchange
interactions are soft, a multiperipheral description in terms of
meson ladders may dominate the physics.  This is the basis for the
two-component pomeron model of Donachie and
Landshoff~\cite{Donnachie:2001xx}.

Consider a collinear frame where the incident momentum $P^+_\pi$
is large and $s = (p_\pi + p_{\rm target})^2 \simeq p^+_\pi
p^-_{\rm target}.$ The matrix element of an exchanged gluon with
momentum $q_i$ between the projectile and an intermediate state
$\ket N$ is dominated by the ``plus current": $\VEV{\pi|j^+(0)\exp
(i {\half q^+_i x^-}-i q_{\perp i} \cdot x_\perp|N}$. Note that
the coherent sum of couplings of an exchanged gluon to the pion
system vanishes when its momentum is small compared to the
characteristic momentum scales in the projectile light-cone
wavefunction: $q^{\perp_i} \Delta x_\perp \ll 1$ and $q^+_i \Delta
x^- \ll 1$.  The destructive interference of the gauge couplings
to the constituents of the projectile follows simply from the fact
that the color charge operator has zero matrix element between
distinct eigenstates of the QCD Hamiltonian: $\VEV{A|Q|B} \equiv
\int d^2x{_\perp} dx^- \VEV{A|j^+(0)|B} = 0$~\cite{BH2}. At high
energies the change in $k^+_i$ of the constituents can be ignored,
so that Fock states of a hadron with small transverse size
interact weakly even in a nuclear target because of their small
dipole moment. This is the basis of ``color transparency'' in
perturbative QCD~\cite{Brodsky:1988xz,Bertsch:1981py}.   To a good
approximation the sum of couplings to the constituents of the
projectile can be represented as a derivative with respect to
transverse momentum. Thus photon exchange measures a weighted sum
of tranverse derivatives $\partial_{k_\perp} \psi_n(x_i,
k_{\perp_i},\lambda_i),$ and two-gluon exchange measures the
second transverse partial derivative~\cite{BHDP}.

3.  The final step is the hadronization of the $n$ constituents of
the projectile Fock state into final state hadrons.  Since $q^+_i$
is small, the number of partons in the initial Fock state and the
final state hadrons are unchanged.  Their coalescence is thus
governed by the convolution of initial and final-state Fock state
wavefunctions.  In the case of states with high $k_\perp$, the
final state will hadronize into jets, each reflecting the
respective $x_i$ of the Fock state constituents. In the case of
higher Fock states with intrinsic sea quarks such as an extra $c
\bar c$ pair (intrinsic charm), one will observe leading $J/\psi$
or open charm hadrons in the projectile fragmentation region; \ie,
the hadron's fragments will tend to have the same rapidity as that
of the projectile.

For example, diffractive multi-jet production in heavy nuclei
provides a novel way to measure the shape of the LC Fock state
wavefunctions and test color transparency.  Consider the reaction
\cite{Bertsch:1981py,Frankfurt:1993it,Frankfurt:2000jm} $\pi A
\rightarrow {\rm Jet}_1 + {\rm Jet}_2 + A^\prime$ at high energy
where the nucleus $A^\prime$ is left intact in its ground state.
The transverse momenta of the jets balance so that $ \vec k_{\perp
i} + \vec k_{\perp 2} = \vec q_\perp < {R^{-1}}_A \ . $ The
light-front longitudinal momentum fractions also need to add to
$x_1+x_2 \sim 1$ so that $\Delta p_L < R^{-1}_A$.  The process can
then occur coherently in the nucleus. Because of color
transparency, the valence wavefunction of the pion with small
impact separation, will penetrate the nucleus with minimal
interactions, diffracting into jet pairs \cite{Bertsch:1981py}.
The $x_1=x$, $x_2=1-x$ dependence of the di-jet distributions will
thus reflect the shape of the pion valence light-front
wavefunction in $x$; similarly, the $\vec k_{\perp 1}- \vec
k_{\perp 2}$ relative transverse momenta of the jets gives key
information on the second derivative of the underlying shape of
the valence pion wavefunction
\cite{Frankfurt:1993it,Frankfurt:2000jm,BHDP}.  The diffractive
nuclear amplitude extrapolated to $t = 0$ should be linear in
nuclear number $A$ if color transparency is correct.  The
integrated diffractive rate should then scale as $A^2/R^2_A \sim
A^{4/3}$.

The results of a diffractive dijet dissociation experiment of this
type E791 at Fermilab using 500 GeV incident pions on nuclear
targets \cite{Aitala:2001hc} appear to be consistent with color
transparency.  The measured longitudinal momentum distribution of
the jets \cite{Aitala:2001hb} is consistent with a pion light-cone
wavefunction of the pion with the shape of the asymptotic
distribution amplitude, $\phi^{\rm asympt}_\pi (x) = \sqrt 3 f_\pi
x(1-x)$.  Data from CLEO \cite{Gronberg:1998fj} for the $\gamma
\gamma^* \rightarrow \pi^0$ transition form factor also favor a
form for the pion distribution amplitude close to the asymptotic
solution to the perturbative QCD evolution equation
\cite{Lepage:1980fj}.

The interpretation of the diffractive dijet processes as measures
of the hadron distribution amplitudes has recently been questioned
by Braun {\em et al.} \cite{Braun:2001ih} and by Chernyak
\cite{Chernyak:2001ph} who have calculated the hard scattering
amplitude for such processes at next-to-leading order.  However,
these analyses neglect the integration over the transverse
momentum of the valence quarks and thus miss the logarithmic
ordering which is required for factorization of the distribution
amplitude and color-filtering in nuclear targets.

As noted above, the diffractive dissociation of a hadron or
nucleus can also occur via the Coulomb dissociation of a beam
particle on an electron beam (\eg\ at HERA or eRHIC) or on the
strong Coulomb field of a heavy nucleus (\eg\ at RHIC or nuclear
collisions at the LHC) \cite{BHDP}.  The amplitude for Coulomb
exchange at small momentum transfer is proportional to the first
derivative $\sum_i e_i {\partial \over \vec k_{T i}} \psi$ of the
light-front wavefunction, summed over the charged constituents.
The Coulomb exchange reactions fall off less fast at high
transverse momentum compared to pomeron exchange reactions since
the light-front wavefunction is effective differentiated twice in
two-gluon exchange reactions.

It will also be interesting to study diffractive tri-jet
production using proton beams $ p A \rightarrow {\rm Jet}_1 + {\rm
Jet}_2 + {\rm Jet}_3 + A^\prime $ to determine the fundamental
shape of the 3-quark structure of the valence light-front
wavefunction of the nucleon at small transverse separation
\cite{Frankfurt:1993it}. For example, consider the Coulomb
dissociation of a high energy proton at HERA.  The proton can
dissociate into three jets corresponding to the three-quark
structure of the valence light-front wavefunction.  We can demand
that the produced hadrons all fall outside an opening angle
$\theta$ in the proton's fragmentation region. Effectively all of
the light-front momentum $\sum_j x_j \simeq 1$ of the proton's
fragments will thus be produced outside an ``exclusion cone".
This then limits the invariant mass of the contributing Fock state
${ M}^2_n > \Lambda^2 = P^{+2} \sin^2\theta/4$ from below, so that
perturbative QCD counting rules can predict the fall-off in the
jet system invariant mass $ M$.  The segmentation of the forward
detector in azimuthal angle $\phi$ can be used to identify
structure and correlations associated with the three-quark
light-front wavefunction \cite{BHDP}. One can use also measure the
dijet structure of real and virtual photons beams $ \gamma^* A
\rightarrow {\rm Jet}_1 + {\rm Jet}_2 + A^\prime $ to measure the
shape of the light-front wavefunction for transversely-polarized
and longitudinally-polarized virtual photons.  Such experiments
will open up a direct window on the amplitude structure of hadrons
at short distances. The light-front formalism is also applicable
to the description of nuclei in terms of their nucleonic and
mesonic degrees of freedom \cite{Miller:2001mi,Miller:2000ta}.
Self-resolving diffractive jet reactions in high energy
electron-nucleus collisions and hadron-nucleus collisions at
moderate momentum transfers can thus be used to resolve the
light-front wavefunctions of nuclei.

\section{Heavy Quark Fluctuations in Diffractive Dissociation}

Since a hadronic wavefunction describes states off of the
light-cone energy shell, there is a finite probability of the
projectile having fluctuations containing extra quark-antiquark
pairs, such as intrinsic strangeness charm, and bottom. In
contrast to the quark pairs arising from gluon splitting,
intrinsic quarks are multiply-connected to the valence quarks and
are thus part of the dynamics of the hadron. Recently Franz,
Polyakov, and Goeke have analyzed the properties of the intrinsic
heavy-quark fluctuations in hadrons using the operator-product
expansion~\cite{Franz:2000ee}. For example, the light-cone
momentum fraction carried by intrinsic heavy quarks in the proton
$x_{Q \bar Q}$ as measured by the $T^{+ + }$ component of the
energy-momentum tensor is related in the heavy-quark limit to the
forward matrix element $\langle p \vert {\hbox{tr}_c}
{(G^{+\alpha} G^{+ \beta} G_{\alpha \beta})/ m_Q^2 }\vert p
\rangle ,$ where $G^{\mu \nu}$ is the gauge field strength tensor.
Diagrammatically, this can be described as a heavy quark loop in
the proton self-energy with four gluons attached to the light,
valence quarks. Since the non-Abelian commutator $[A_\alpha,
A_\beta]$ is involved, the heavy quark pairs in the proton
wavefunction are necessarily in a color-octet state. It follows
from dimensional analysis that the momentum fraction carried by
the $Q\bar Q$ pair scales as $k^2_\perp / m^2_Q$ where $k_\perp$
is the typical momentum in the hadron wave function.  [In
contrast, in the case of Abelian theories, the contribution of an
intrinsic, heavy lepton pair to the bound state's structure first
appears in ${ O}(1/m_L^4)$.  One relevant operator corresponds to
the Born-Infeld $(F_{\mu\nu})^4$ light-by-light scattering
insertion, and the momentum fraction of heavy leptons in an atom
scales as $k^4_\perp / m_L^4$.]

Intrinsic charm can be materialized by diffractive dissociation
into open or hidden charm states such as $p p \to J/\psi X p',
\Lambda_c X p'$. At HERA one can measure intrinsic charm in the
proton by Coulomb dissociation: $p e \to \Lambda_C X e',$ and
$J/\psi X e'.$  Since the intrinsic heavy quarks tend to have the
same rapidity as that of the projectile, they are produced at
large $x_F$ in the beam fragmentation region. The charm structure
function measured by the EMC group shows an excess at large
$x_{bj}$, indicating a probability of order $1\%$ for intrinsic
charm in the proton~\cite{Harris:1996jx}. The presence of
intrinsic charm in light-mesons provides an explanation for the
puzzle of the large $J/\psi \to \rho\pi$ branching ratio and
suppressed $\psi^\prime \to \rho\pi$ decay~\cite{Brodsky:1997fj}.
The presence of intrinsic charm quarks in the $B$ wave function
provides new mechanisms for $B$ decays.  For example, Chang and
Hou have considered the production of final states with three
charmed quarks such as $B \to J/\psi D \pi$ and $B \to J/\psi
D^*$~\cite{Chang:2001iy}; these final states are difficult to
realize in the valence model, yet they occur naturally when the
$b$ quark of the intrinsic charm Fock state $\ket{ b \bar u c \bar
c}$ decays via $b \to c \bar u d$.  In fact, the $J/\psi$ spectrum
for inclusive $B \to J/\psi X$ decays measured by CLEO and Belle
shows a distinct enhancement at the low $J/\psi$ momentum where
such decays would kinematically occur.  Alternatively, this excess
could reflect the opening of baryonic channels such as $B \to
J/\psi \bar p \Lambda$~\cite{Brodsky:1997yr}. Recently, Susan
Gardner and I have shown that the presence of intrinsic charm in
the hadrons' light-cone wave functions, even at a few percent
level, provides new, competitive decay mechanisms for $B$ decays
which are nominally CKM-suppressed~\cite{Brodsky:2001yt}.  For
example, the weak decays of the $B$-meson to two-body exclusive
states consisting of strange plus light hadrons, such as $B \to
\pi K$, are expected to be dominated by penguin contributions
since the tree-level $b\to s u{\overline u}$ decay is CKM
suppressed. However, higher Fock states in the $B$ wave function
containing charm quark pairs can mediate the decay via a
CKM-favored $b\to s c{\overline c}$ tree-level transition. Such
intrinsic charm contributions can be phenomenologically
significant.  Since they mimic the amplitude structure of
``charming'' penguin contributions~\cite{Ciuchini:2001gv},
charming penguins need not be penguins at
all~\cite{Brodsky:2001yt}.

\section{Calculating and Modeling Light-Cone Wavefunctions}

The discretized light-cone quantization method~\cite{Pauli:1985pv}
is a powerful technique for finding the non-perturbative solutions
of quantum field theories.  The basic method is to diagonalize the
light-cone Hamiltonian in a light-cone Fock basis defined using
periodic boundary conditions in $x^-$ and $x_\perp$.  The method
preserves the frame-independence of the front form.  The DLCQ
method is now used extensively to solve one-space and one-time
theories, including supersymmetric theories. New applications of
DLCQ to supersymmetric quantum field theories and specific tests
of the Maldacena conjecture have recently been given by Pinsky and
Trittman.  There has been progress in systematically developing
the computation and renormalization methods needed to make DLCQ
viable for QCD in physical spacetime. For example, John Hiller,
Gary McCartor and I \cite{Brodsky:2001ja} have shown how DLCQ can
be used to solve 3+1 theories despite the large numbers of degrees
of freedom needed to enumerate the Fock basis.  A key feature of
our work, is the introduction of Pauli Villars fields in order to
regulate the UV divergences and perform renormalization while
preserving the frame-independence of the theory. A review of DLCQ
and its applications is given in Ref. \cite{Brodsky:1998de}.
There has also been important progress using the transverse
lattice, essentially a combination of DLCQ in 1+1 dimensions
together with a lattice in the transverse dimensions.

Even without explicit solutions, many features of the light-cone
wavefunctions follow from general arguments. Light-cone
wavefunctions satisfy the equation of motion:
$$ H^{QCD}_{LC} \ket{\Psi} = (H^{0}_{LC} + V_{LC} )\ket{\Psi} = M^2
\ket{\Psi},$$ which has the Heisenberg matrix form in Fock space:
$$M^2 - \sum_{i=1}^n{m_{\perp i}^2\over x_i} \psi_n =
\sum_{n'}\int \VEV{n|V|n'} \psi_{n'}$$
where the convolution and sum is understood over the Fock number,
transverse momenta, plus momenta and helicity of the intermediate
states.  Here $m^2_\perp = m^2 + k^2_\perp.$ Thus, in general,
every light-cone Fock wavefunction has the form:
$$\psi_n={\Gamma_n\over M^2-\sum_{i=1}^n{m_{\perp i}^2\over x_i}}$$
where $\Gamma_n = \sum_{n'}\int V_{n {n'}} \psi_n$. The main
dynamical dependence of a light-cone wavefunction away from the
extrema is controlled by its light-cone energy denominator.  The
maximum of the wavefunction occurs when the invariant mass of the
partons is minimal; \ie, when all particles have equal rapidity
and are all at rest in the rest frame. In fact, Dae  Sung Hwang
and I \cite{BH} have noted that one can rewrite the wavefunction
in the form:
$$\psi_n= {\Gamma_n\over M^2
[\sum_{i=1}^n {(x_i-{\hat x}_i)^2\over x_i} + \delta^2]}$$
where $x_i = {\hat x}_i\equiv{m_{\perp i}/ \sum_{i=1}^n m_{\perp
i}}$ is the condition for minimal rapidity differences of the
constituents.  The key parameter is $ M^2-\sum_{i=1}^n{m_{\perp
i}^2/ {\hat x}_i}\equiv -M^2\delta^2.$ We can also interpret
$\delta^2 \simeq 2 \epsilon / M $ where $ \epsilon = \sum_{i=1}^n
m_{\perp i}-M $ is the effective binding energy. This form shows
that the wavefunction is a quadratic form around its maximum, and
that the width of the distribution in $(x_i - \hat x_i)^2$ (where
the wavefunction falls to half of its maximum) is controlled by
$x_i \delta^2$ and the transverse momenta $k_{\perp_i}$.  Note
also that the heaviest particles tend to have the largest $\hat
x_i,$ and thus the largest momentum fraction of the particles in
the Fock state, a feature familiar from the intrinsic charm model.
For example, the $b$ quark has the largest momentum fraction at
small $k_\perp$ in the $B$ meson's valence light-cone
wavefunction,, but the distribution spreads out to an
asymptotically symmetric distribution around $x_b \sim 1/2$ when
$k_\perp >> m^2_b.$

We can also discern some general properties of the numerator of
the light-cone wavefunctions. $\Gamma_n(x_i, k_{\perp i},
\lambda_i)$. The transverse momentum dependence of $\Gamma_n$
guarantees $J_z$ conservation for each Fock state: Each light-cone
Fock wavefunction satisfies conservation of the $z$ projection of
angular momentum: $ J^z = \sum^n_{i=1} S^z_i + \sum^{n-1}_{j=1}
l^z_j \ . $ The sum over $s^z_i$ represents the contribution of
the intrinsic spins of the $n$ Fock state constituents.  The sum
over orbital angular momenta $l^z_j = -{\mathrm i}
(k^1_j\frac{\partial}{\partial k^2_j}
-k^2_j\frac{\partial}{\partial k^1_j})$ derives from the $n-1$
relative momenta.  This excludes the contribution to the orbital
angular momentum due to the motion of the center of mass, which is
not an intrinsic property of the hadron \cite{Brodsky:2001ii}. For
example, one of the three light-cone Fock wavefunctions of a $J_z
= +1/2$ lepton in QED perturbation theory is $
\psi^{\uparrow}_{+\frac{1}{2}\, +1} (x,{\vec
k}_{\perp})=-{\sqrt{2}} \frac{(-k^1+{\mathrm i} k^2)}{x(1-x)}\,
\varphi \ ,$ where $ \varphi=\varphi (x,{\vec k}_{\perp})=\frac{
e/\sqrt{1-x}}{M^2-({\vec k}_{\perp}^2+m^2)/x-({\vec
k}_{\perp}^2+\lambda^2)/(1-x)}\ . $ The orbital angular momentum
projection in this case is $\ell^z = -1.$ The spin structure
indicated by perturbative theory provides a template for the
numerator structure of the light-cone wavefunctions even for
composite systems. The structure of the electron's Fock state in
perturbative QED shows that it is natural to have a negative
contribution from relative orbital angular momentum which balances
the $S_z$ of its photon constituents. We can also expect a
significant orbital contribution to the proton's $J_z$ since
gluons carry roughly half of the proton's momentum, thus providing
insight into the ``spin crisis" in QCD.

The fall-off the light-cone wavefunctions at large $k_\perp$ and
$x \to 1$ is dictated by QCD perturbation theory since the state
is far-off the light-cone energy shell.  This leads to counting
rule behavior for the quark and gluon distributions at $x \to 1$.
Notice that $x\to 1$ corresponds to $k^z \to -\infty$ for any
constituent with nonzero mass or transverse momentum.

The above discussion suggests that an approximate form for the
hadron light-cone wavefunctions might be constructed through
variational principles and by minimizing the expectation value of
$H^{QCD}_{LC}.$

\section{Structure Functions are Not Parton Distributions}

Ever since the earliest days of the parton model, it has been
assumed that the leading-twist structure functions $F_i(x,Q^2)$
measured in deep inelastic lepton scattering are determined by the
{\it probability} distribution of quarks and gluons as determined
by the light-cone wavefunctions of the target.  For example, the
quark distribution is
$$
{ P}_{\qu/N}(x_B,Q^2)= \sum_n \int^{k_{i\perp}^2<Q^2}\left[
\prod_i\, dx_i\, d^2k_{\perp i}\right] |\psi_n(x_i,k_{\perp i})|^2
\sum_{j=q} \delta(x_B-x_j).
$$
The identification of structure functions with the square of
light-cone wavefunctions is usually made in LC gauge $n\cdot A =
A^+=0$, where the path-ordered exponential in the operator product
for the forward virtual Compton amplitude apparently reduces to
unity. Thus the deep inelastic lepton scattering cross section
(DIS) appears to be fully determined by the probability
distribution of partons in the target. However, Paul Hoyer, Nils
Marchal, Stephane Peigne, Francesco Sannino, and I have recently
shown that the leading-twist contribution to DIS is affected by
diffractive rescattering of a quark in the target, a coherent
effect which is not included in the light-cone wavefunctions, even
in light-cone gauge.  The distinction between structure functions
and parton probabilities is already implied by the Glauber-Gribov
picture of nuclear
shadowing~\cite{Gribov:1969jf,Brodsky:1969iz,Brodsky:1990qz,Piller:2000wx}.
In this framework shadowing arises from interference between
complex rescattering amplitudes involving on-shell intermediate
states, as in Fig. 1.  In contrast, the wave function of a stable
target is strictly real since it does not have on energy-shell
configurations.  A probabilistic interpretation of the DIS cross
section is thus precluded.

It is well-known that in Feynman and other covariant gauges one
has to evaluate the corrections to the ``handbag" diagram due to
the final state interactions of the struck quark (the line
carrying momentum $p_1$ in Fig. 2) with the gauge field of the
target.  In light-cone gauge, this effect also involves
rescattering of a spectator quark, the $p_2$ line in Fig. 2.  The
light-cone gauge is singular -- in particular, the gluon
propagator $ d_{LC}^{\mu\nu}(k) =
\frac{i}{k^2+\ieps}\left[-g^{\mu\nu}+\frac{n^\mu k^\nu+ k^\mu
n^\nu}{n\cdot k}\right] \label{lcprop} $ has a pole at $k^+ = 0$
which requires an analytic prescription.  In final-state
scattering involving on-shell intermediate states, the exchanged
momentum $k^+$ is of \order{1/\nu} in the target rest frame, which
enhances the second term in the propagator.  This enhancement
allows rescattering to contribute at leading twist even in LC
gauge.

\begin{figure}[htb]
\centering
\includegraphics[width=5in]{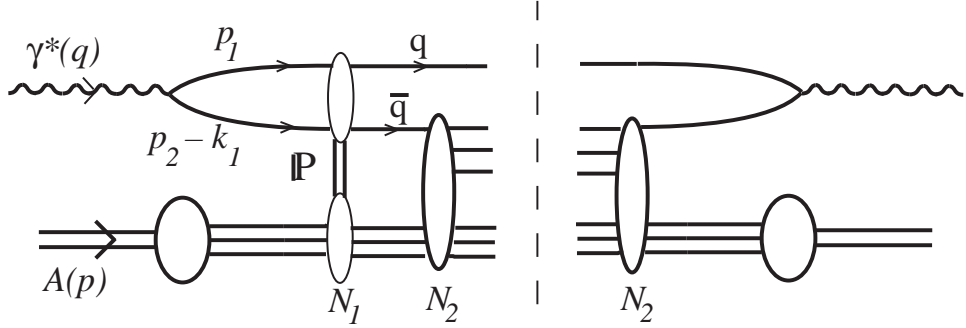}
\caption[*]{Glauber-Gribov shadowing involves interference between
rescattering amplitudes. \label{brodsky2}}
\end{figure}

%%%%%%%%%%%%%%%%%%%%%%%%%

The issues involving final state interactions even occur in the
simple framework of abelian gauge theory with scalar quarks.
Consider a frame with $q^+ < 0$. We can then distinguish FSI from
ISI using LC time-ordered perturbation
theory~\cite{Lepage:1980fj}. Figure 1 illustrates two LCPTH
diagrams which contribute to the forward $\gamma^* T \to \gamma^*
T$ amplitude, where the target $T$ is taken to be a single quark.
In the aligned jet kinematics the virtual photon fluctuates into a
\qu\qb\ pair with limited transverse momentum, and the (struck)
quark takes nearly all the longitudinal momentum of the photon.
The initial \qu\ and \qb\ momenta are denoted $p_1$ and $p_2-k_1$,
respectively,

%%%%%%%%%%%%%%%%%%%%%%%%%
\begin{figure}[htb]
\includegraphics[width=5in]{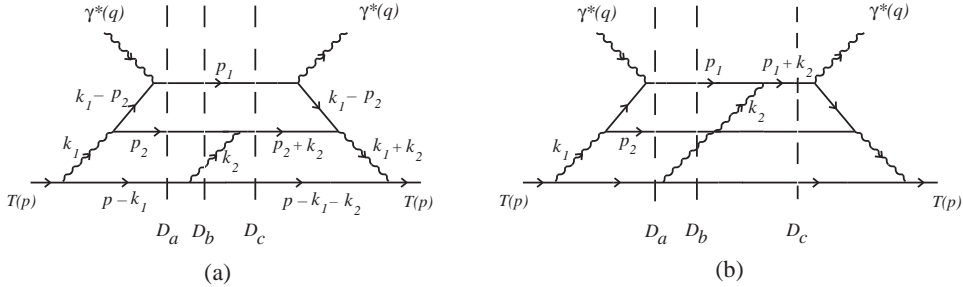}
\caption[*]{Two types of final state interactions.  (a) Scattering
of the antiquark ($p_2$ line), which in the aligned jet kinematics
is part of the target dynamics.  (b) Scattering of the current
quark ($p_1$ line).  For each LC time-ordered diagram, the
potentially on-shell intermediate states -- corresponding to the
zeroes of the denominators $D_a, D_b, D_c$ -- are denoted by
dashed lines.} \label{brodsky1}
\end{figure}
%%%%%%%%%%%%%%%%%%%%%%%%%

The calculation of the rescattering effect of DIS in Feynman and
light-cone gauge through three loops is given in detail in
Ref.~\cite{Brodsky:2001ue}.  The result can be resummed and is
most easily expressed in eikonal form in terms of transverse
distances $r_\perp, R_\perp$ conjugate to $p_{2\perp}, k_\perp$.
The deep inelastic cross section can be expressed as \beq
Q^4\frac{d\sigma}{dQ^2\, dx_B} =
\frac{\alpha}{16\pi^2}\frac{1-y}{y^2} \frac{1}{2M\nu} \int
\frac{dp_2^-}{p_2^-}\,d^2\rvec_\perp\, d^2\Rvec_\perp\, |\tilde
M|^2 \label{transcross} \eeq where \beq |\tilde{
M}(p_2^-,\rvec_\perp, \Rvec_\perp)| = \left|\frac{\sin \left[g^2\,
W(\rvec_\perp, \Rvec_\perp)/2\right]}{g^2\, W(\rvec_\perp,
\Rvec_\perp)/2} \tilde{A}(p_2^-,\rvec_\perp, \Rvec_\perp)\right|
\label{Interference} \eeq is the resummed result.  The Born
amplitude is \beq \tilde A(p_2^-,\rvec_\perp, \Rvec_\perp) = 2eg^2
M Q p_2^-\, V(m_\pl r_\perp) W(\rvec_\perp, \Rvec_\perp)
\label{Atildeexpr} \eeq where $ m_\pl^2 = p_2^-Mx_B + m^2
\label{mplus}$ and \beq V(m\, r_\perp) \equiv \int
\frac{d^2\pvec_\perp}{(2\pi)^2}
\frac{e^{i\rvec_\perp\cdot\pvec_{\perp}}}{p_\perp^2+m^2} =
\frac{1}{2\pi}K_0(m\,r_\perp) \label{Vexpr} \eeq The rescattering
effect of the dipole of the $q \bar q$ is controlled by \beq
W(\rvec_\perp, \Rvec_\perp) \equiv \int
\frac{d^2\kvec_\perp}{(2\pi)^2}
\frac{1-e^{i\rvec_\perp\cdot\kvec_{\perp}}}{k_\perp^2}
e^{i\Rvec_\perp\cdot\kvec_{\perp}} = \frac{1}{2\pi}
\log\left(\frac{|\Rvec_\perp+\rvec_\perp|}{R_\perp} \right).
\label{Wexpr} \eeq The fact that the coefficient of $\tilde A$ in
\eq{Interference} is less than unity for all $\rvec_\perp,
\Rvec_\perp$ shows that the rescattering corrections reduce the
cross section.  It is the analog of nuclear shadowing in our
model.

We have also found the same result for the deep inelastic cross
sections in light-cone gauge.  Three prescriptions for defining
the propagator pole at $k^+ =0$ have been used in the literature:
\beq \label{prescriptions} \frac{1}{k_i^+} \rightarrow
\left[\frac{1}{k_i^+} \right]_{\eta_i} = \left\{
\begin{array}{cc}
k_i^+\left[(k_i^+ -i\eta_i)(k_i^+ +i\eta_i)\right]^{-1} & ({\rm PV}) \\
\left[k_i^+ -i\eta_i\right]^{-1} & ({\rm K}) \\
\left[k_i^+ -i\eta_i \epsilon(k_i^-)\right]^{-1} & ({\rm ML})
\end{array} \right.
\eeq the principal-value, Kovchegov~\cite{Kovchegov:1997pc}, and
Mandelstam-Leibbrandt~\cite{Leibbrandt:1987qv} prescriptions. The
`sign function' is denoted $\epsilon(x)=\Theta(x)-\Theta(-x)$.
With the PV prescription we have $ I_{\eta} = \int dk_2^+
\left[\frac{1}{k_2^+} \right]_{\eta_2} = 0. $ Since an individual
diagram may contain pole terms $\sim 1/k_i^+$, its value can
depend on the prescription used for light-cone gauge. However, the
$k_i^+=0$ poles cancel when all diagrams are added;  the net is
thus prescription-independent, and it agrees with the Feynman
gauge result. It is interesting to note that the diagrams
involving rescattering of the struck quark $p_1$ do not contribute
to the leading-twist structure functions if we use the Kovchegov
prescription to define the light-cone gauge.  In other
prescriptions for light-cone gauge the rescattering of the struck
quark line $p_1$ leads to an infrared divergent phase factor $\exp
i\phi$: \beq \phi = g^2 \, \frac{I_{\eta}-1}{4 \pi} \, K_0(\lambda
R_{\perp}) + {{O}}(g^6) \eeq where $\lambda$ is an infrared
regulator, and $I_{\eta}= 1$ in the $K$ prescription. The phase is
exactly compensated by an equal and opposite phase from
final-state interactions of line $p_2$. This irrelevant change of
phase can be understood by the fact that the different
prescriptions are related by a residual gauge transformation
proportional to $\delta(k^+)$ which leaves the light-cone gauge
$A^+ = 0$ condition unaffected.

Diffractive contributions which leave the target intact thus
contribute at leading twist to deep inelastic scattering.  These
contributions do not resolve the quark structure of the target,
and thus they are contributions to structure functions which are
not parton probabilities. More generally, the rescattering
contributions shadow and modify the observed inelastic
contributions to DIS.

Our analysis in the $K$ prescription for light-cone gauge
resembles the ``covariant parton model" of Landshoff, Polkinghorne
and Short~\cite{Landshoff:1971ff,Brodsky:1973hm} when interpreted
in the target rest frame.  In this description of small $x$ DIS,
the virtual photon with positive $q^+$ first splits into the pair
$p_1$ and $p_2$.  The aligned quark $p_1$ has no final state
interactions.  However, the antiquark line $p_2$ can interact in
the target with an effective energy $\hat s \propto {k_\perp^2/x}$
while staying close to its mass shell.  Thus at small $x$ and
large $\hat s$, the antiquark $p_2$ line can first multiple
scatter in the target via pomeron and Reggeon exchange, and then
it can finally scatter inelastically or be annihilated. The DIS
cross section can thus be written as an integral of the
$\sigma_{\bar q p \to X}$ cross section over the $p_2$ virtuality.
In this way, the shadowing of the antiquark in the nucleus
$\sigma_{\bar q A \to X}$ cross section yields the nuclear
shadowing of DIS~\cite{Brodsky:1990qz}.  Our analysis, when
interpreted in frames with $q^+ > 0,$ also supports the color
dipole description of deep inelastic lepton scattering at small
$x$.  Even in the case of the aligned jet configurations, one can
understand DIS as due to the coherent color gauge interactions of
the incoming quark-pair state of the photon interacting first
coherently and finally incoherently in the target.

\section{Acknowledgment} I thank Professor Vojtech Kundrat for organizing this
outstanding meeting.  Much of the new work reported here was done
in collaboration with others, especially,  Susan Gardner, Dae Sung
Hwang, Paul Hoyer, Nils Marchal, Stephane Peigne, and Francesco
Sannino. This work was supported by the Department of Energy under
contract number DE-AC03-76SF00515.


\begin{thebibliography}{00}

%\cite{Brodsky:1999hn}
\bibitem{Brodsky:1999hn}
S.~J.~Brodsky and D.~S.~Hwang,
%``Exact light-cone wavefunction representation of matrix elements of
% electroweak currents,''
Nucl.\ Phys.\ B {\bf 543}, 239 (1999) [hep-ph/9806358].
%%CITATION = HEP-PH 9806358;%%

\bibitem{Brodsky:1999kn}
S.~J.~Brodsky, V.~S.~Fadin, V.~T.~Kim, L.~N.~Lipatov and
G.~B.~Pivovarov,
%``The {QCD} pomeron with optimal renormalization,''
JETP Lett.\  {\bf 70}, 155 (1999) [hep-ph/9901229].
%%CITATION = HEP-PH 9901229;%%


\bibitem{Donnachie:2001xx}
A.~Donnachie and P.~V.~Landshoff,
%``New data and the hard pomeron,''
hep-ph/0105088.
%%CITATION = HEP-PH 0105088;%%



\bibitem{BH2}
S.~J.~Brodsky and D.-S. Hwang, in preparation.

\bibitem{Brodsky:1988xz} S.~J.~Brodsky and
A.~H.~Mueller,
%``Using Nuclei To Probe Hadronization In QCD,''
Phys.\ Lett.\ B {\bf 206}, 685 (1988).
%%CITATION = PHLTA,B206,685;%%



%\cite{Bertsch:1981py}
\bibitem{Bertsch:1981py}
G.~Bertsch, S.~J.~Brodsky, A.~S.~Goldhaber and J.~F.~Gunion,
%``Diffractive Excitation In QCD,''
Phys.\ Rev.\ Lett.\  {\bf 47}, 297 (1981).
%%CITATION = PRLTA,47,297;%%

\bibitem{BHDP}
S. Brodsky, P.Hoyer, Markus Diehl,  S. Peigne, and W. Sch\"afer,
(in preparation).


%\cite{Frankfurt:1993it}
\bibitem{Frankfurt:1993it}
L.~Frankfurt, G.~A.~Miller and M.~Strikman,
%``Coherent nuclear diffractive production of mini - jets: Illuminating
% color transparency,''
Phys.\ Lett.\ B {\bf 304}, 1 (1993) [hep-ph/9305228].
%%CITATION = HEP-PH 9305228;%%

%\cite{Frankfurt:2000jm}
\bibitem{Frankfurt:2000jm}
L.~Frankfurt, G.~A.~Miller and M.~Strikman,
%``Coherent QCD phenomena in the coherent pion nucleon and pion nucleus
%production of two jets at high relative momenta,''
hep-ph/0010297.
%%CITATION = HEP-PH 0010297;%%

%\cite{Aitala:2001hc}
\bibitem{Aitala:2001hc}
E.~M.~Aitala {\it et al.}  [E791 Collaboration],
%``Observation of color-transparency in diffractive dissociation of pions,''
Phys.\ Rev.\ Lett.\  {\bf 86}, 4773 (2001) [hep-ex/0010044].
%%CITATION = HEP-EX 0010044;%%

%\cite{Aitala:2001hb}
\bibitem{Aitala:2001hb}
E.~M.~Aitala {\it et al.}  [E791 Collaboration],
%``Direct measurement of the pion valence quark momentum distribution, the
%pion light-cone wave function squared,''
Phys.\ Rev.\ Lett.\  {\bf 86}, 4768 (2001) [hep-ex/0010043].
%%CITATION = HEP-EX 0010043;%%

%\cite{Gronberg:1998fj}
\bibitem{Gronberg:1998fj}
J.~Gronberg {\it et al.}  [CLEO Collaboration],
%``Measurements of the meson photon transition form factors of light
% pseudoscalar mesons at large momentum transfer,''
Phys.\ Rev.\ D {\bf 57}, 33 (1998) [hep-ex/9707031].
%%CITATION = HEP-EX 9707031;%%

%\cite{Lepage:1980fj}
\bibitem{Lepage:1980fj}
G.~P.~Lepage and S.~J.~Brodsky,
%``Exclusive Processes In Perturbative Quantum Chromodynamics,''
Phys.\ Rev.\ D {\bf 22}, 2157 (1980).
%%CITATION = PHRVA,D22,2157;%%%\cite{Miller:2000ta}

%
\bibitem{Braun:2001ih}
V.~M.~Braun, D.~Y.~Ivanov, A.~Schafer and L.~Szymanowski,
%``QCD factorization for the pion diffractive dissociation to two jets,''
\emph{Phys.\ Lett.\ B} {\bf 509}, 43 (2001) [hep-ph/0103275].
%%CITATION = HEP-PH 0103275;%%

%
\bibitem{Chernyak:2001ph}
V.~Chernyak,
%``Does the E791 experiment have measured the pion wave function profile?,''
[hep-ph/0103295].
%%CITATION = HEP-PH 0103295;%%

%\cite{Miller:2001mi}
\bibitem{Miller:2001mi}
G.~A.~Miller,
%``Infinite nuclear matter on the light front: A modern approach to
% Brueckner theory,''
Int.\ J.\ Mod.\ Phys.\ B {\bf 15}, 1551 (2001) [nucl-th/9910053].
%%CITATION = NUCL-TH 9910053;%%

\bibitem{Miller:2000ta}
G.~A.~Miller, S.~J.~Brodsky and M.~Karliner,
%``Coherent contributions of nuclear mesons to electroproduction and the
% HERMES effect,''
Phys.\ Lett.\ B {\bf 481}, 245 (2000) [hep-ph/0002156].
%%CITATION = HEP-PH 0002156;%%

%\cite{Franz:2000ee}
\bibitem{Franz:2000ee}
M.~Franz,~V.~Polyakov and K.~Goeke,
%``Heavy quark mass expansion and intrinsic charm in light hadrons,''
Phys.\ Rev.\ D {\bf 62}, 074024 (2000) [hep-ph/0002240].
%%CITATION = HEP-PH 0002240;%%

%\cite{Harris:1996jx}
\bibitem{Harris:1996jx}
B.~W.~Harris, J.~Smith and R.~Vogt,
%``Reanalysis of the EMC charm production data with extrinsic and intrinsic
% charm at NLO,''
Nucl.\ Phys.\ B {\bf 461}, 181 (1996) [hep-ph/9508403].
%%CITATION = HEP-PH 9508403;%%

%\cite{Brodsky:1997fj}
\bibitem{Brodsky:1997fj}
S.~J.~Brodsky and M.~Karliner,
%``Intrinsic charm of vector mesons: A possible solution of the *rho pi
% puzzle*,''
Phys.\ Rev.\ Lett.\  {\bf 78}, 4682 (1997) [hep-ph/9704379].
%%CITATION = HEP-PH 9704379;%%

%\cite{Chang:2001iy}
\bibitem{Chang:2001iy}
C.~V.~Chang and W.~Hou,
%``anti-B $\to$ J/psi D (pi) as smoking gun evidence for intrinsic charm in
% B meson,''
hep-ph/0101162.
%%CITATION = HEP-PH 0101162;%%


%\cite{Brodsky:1997yr}
\bibitem{Brodsky:1997yr}
S.~J.~Brodsky and F.~S.~Navarra,
%``Looking for exotic multiquark states in nonleptonic B decays,''
Phys.\ Lett.\ B {\bf 411}, 152 (1997) [hep-ph/9704348].
%%CITATION = HEP-PH 9704348;%%

%\cite{Brodsky:2001yt}
\bibitem{Brodsky:2001yt}
S.~J.~Brodsky and S.~Gardner,
%``Evading the CKM hierarchy,''
hep-ph/0108121.
%%CITATION = HEP-PH 0108121;%%

\bibitem{Ciuchini:2001gv}
M.~Ciuchini, E.~Franco, G.~Martinelli, M.~Pierini and
L.~Silvestrini,
%``Charming penguins strike back,''
Phys.\ Lett.\ B {\bf 515}, 33 (2001) [hep-ph/0104126].
%%CITATION = HEP-PH 0104126;%%

\bibitem{Pauli:1985pv}
H.~C.~Pauli and S.~J.~Brodsky,
%``Solving Field Theory In One Space One Time Dimension,''
Phys.\ Rev.\ D {\bf 32}, 1993 (1985).
%%CITATION = PHRVA,D32,1993;%%

%\cite{Brodsky:2001ja}
\bibitem{Brodsky:2001ja}
S.~J.~Brodsky, J.~R.~Hiller and G.~McCartor,
%``Application of Pauli-Villars regularization and discretized light-cone
% quantization to a single-fermion truncation of Yukawa theory,''
hep-ph/0107038.
%%CITATION = HEP-PH 0107038;%%

%\cite{Brodsky:1998de}
\bibitem{Brodsky:1998de}
S.~J.~Brodsky, H.~Pauli and S.~S.~Pinsky,
%``Quantum chromodynamics and other field theories on the light cone,''
Phys.\ Rept.\  {\bf 301}, 299 (1998) [hep-ph/9705477].
%%CITATION = HEP-PH 9705477;%%

\bibitem{BH} S. J. Brodsky and D. S. Hwang, in preparation.


%\cite{Brodsky:2001ii}
\bibitem{Brodsky:2001ii}
S.~J.~Brodsky, D.~S.~Hwang, B.~Ma and I.~Schmidt,
%``Light-cone representation of the spin
% and orbital angular momentum of relativistic composite systems,''
Nucl.\ Phys.\ B {\bf 593}, 311 (2001) [hep-th/0003082].
%%CITATION = HEP-TH 0003082;%%


%\cite{Gribov:1969jf}
\bibitem{Gribov:1969jf}
V.~N.~Gribov,
%``Glauber Corrections And The Interaction Between High-Energy Hadrons And
% Nuclei,''
Sov.\ Phys.\ JETP {\bf 29}, 483 (1969) [Zh.\ Eksp.\ Teor.\ Fiz.\
{\bf 56}, 892 (1969)].
%%CITATION = SPHJA,29,483;%%


%\cite{Brodsky:1969iz}
\bibitem{Brodsky:1969iz}
S.~J.~Brodsky and J.~Pumplin,
%``Photon - Nucleus Total Cross-Sections,''
Phys.\ Rev.\  {\bf 182}, 1794 (1969).
%%CITATION = PHRVA,182,1794;%%


%\cite{Brodsky:1990qz}
\bibitem{Brodsky:1990qz}
S.~J.~Brodsky and H.~J.~Lu,
%``Shadowing And Antishadowing Of Nuclear Structure Functions,''
Phys.\ Rev.\ Lett.\  {\bf 64}, 1342 (1990).
%%CITATION = PRLTA,64,1342;%%




%\cite{Piller:2000wx}
\bibitem{Piller:2000wx}
G.~Piller and W.~Weise,
%``Nuclear deep-inelastic lepton scattering and coherence phenomena,''
Phys.\ Rept.\  {\bf 330}, 1 (2000) [hep-ph/9908230].
%%CITATION = HEP-PH 9908230;%%


%\cite{Brodsky:1980pb}
%\bibitem{Brodsky:1980pb}
%S.~J.~Brodsky, P.~Hoyer, C.~Peterson and N.~Sakai,
%%``The Intrinsic Charm Of The Proton,''
%Phys.\ Lett.\ B {\bf 93}, 451 (1980).
%%%CITATION = PHLTA,B93,451;%%







%\cite{Brodsky:2001ue}
\bibitem{Brodsky:2001ue}
S.~J.~Brodsky, P.~Hoyer, N.~Marchal, S.~Peigne and F.~Sannino,
%``Structure functions are not parton probabilities,''
hep-ph/0104291.
%%CITATION = HEP-PH 0104291;%%




%\cite{Kovchegov:1997pc}
\bibitem{Kovchegov:1997pc}
Y.~V.~Kovchegov,
%``Quantum structure of the non-Abelian Weizsaecker-Williams field for a
% very large nucleus,''
Phys.\ Rev.\ D {\bf 55}, 5445 (1997) [hep-ph/9701229].
%%CITATION = HEP-PH 9701229;%%

%\cite{Leibbrandt:1987qv}
\bibitem{Leibbrandt:1987qv}
G.~Leibbrandt,
%``Introduction To Noncovariant Gauges,''
Rev.\ Mod.\ Phys.\  {\bf 59}, 1067 (1987).
%%CITATION = RMPHA,59,1067;%%


\bibitem{Landshoff:1971ff}
P.~V.~Landshoff, J.~C.~Polkinghorne and R.~D.~Short,
%``A Nonperturbative Parton Model Of Current Interactions,''
Nucl.\ Phys.\ B {\bf 28}, 225 (1971).
%%CITATION = NUPHA,B28,225;%%


%\cite{Brodsky:1973hm}
\bibitem{Brodsky:1973hm}
S.~J.~Brodsky, F.~E.~Close and J.~F.~Gunion,
%``A Gauge - Invariant Scaling Model Of Current Interactions With Regge
% Behavior And Finite Fixed Pole Sum Rules,''
Phys.\ Rev.\ D {\bf 8}, 3678 (1973).
%%CITATION = PHRVA,D8,3678;%%



\end{thebibliography}
\end{document}